\documentclass[submission,copyright,creativecommons]{eptcs}
\providecommand{\event}{HAS 2013} % Name of the event you are submitting to
\usepackage{amssymb}
\usepackage{graphicx}
\usepackage{amsmath}
\usepackage{stmaryrd}                       % Enables \llbracket
\usepackage{dsfont}                         % Enables double stroke fonts
\usepackage{bm}
\usepackage{mathdots}                       % Enables \iddots
\usepackage{xspace}
\usepackage[draft,inline,nomargin]{fixme}
\usepackage{graphicx}
\usepackage{color}
\usepackage{amsthm}

\newtheorem{example}{Example}

\theoremstyle{remark}

\usepackage{listings}
\title{Combinatorial Abstractions of Dynamical Systems}
\author{Rafael Wisniewski\thanks{This work was supported by MT-LAB, a VKR Centre of Excellence for the Modeling of Information Technology.}
\institute{Section of Automation \& Control\\
Aalborg University, Denmark}
\email{raf@es.aau.dk}
}
\def\titlerunning{Completeness of Lyapunov Abstraction}
\def\authorrunning{R. Wisniewski}
\begin{document}
\maketitle

\section{Extended Abstract}

%%%%%
Formal verification has been successfully developed in computer science for verifying combinatorial classes of models and specifications \cite{Clarke:1986:AVF:5397.5399}. In like manner, formal verification methods have been developed for dynamical systems \cite{tabuada2009verification}. However, the verification of system properties, such as safety, is based on reachability calculations, which are the sources of insurmountable complexity. This talk addresses indirect verification methods, which are based on abstracting the dynamical systems by models of reduced complexity and preserving central properties of the original systems. 

Specifically, in this talk, I consider a dynamical system $\mathcal C = (M,\xi)$, where $M$ is the state space - a closed manifold, and $\xi$ is a smooth vector field on $M$. 

We denote a flow line
of $\xi$ by $\phi_x(t) \equiv \phi_x^{\xi} (t)$, that is
\[\frac{d}{dt} \phi_x (t) = \xi\left({\phi_x(t)}\right) \hbox{ with } \phi_x(0) =
x.\] The manifold $M$ is compact; thus, the vector field $\xi$ generates a
1-parameter group $\phi_t: M \rightarrow M,~t \in \Bbb R,$ of
diffeomorphisms. The smooth flow map $\phi: \Bbb R \times M
\rightarrow M$ is related to $\phi_t$ in the following way
\[\phi(t,x) \equiv \phi_t(x) \equiv \phi_x(t).\]

We will examine examples of candidates for the combinatorial system $\mathcal D$ that mirrors the behaviour of $C$. For now, the combinatorial system $\mathcal D$ is a pair $(Z,\Phi)$ consisting of a finite set $Z$, and a function $\Phi: \mathbb R \times Z \to 2^Z$, where $2^Z$ denotes  the power set of $Z$.  We think about $Z$ as a discrete state space and about $\Phi$ as a discrete flow map. Subsequently, we will discuss methods of converting the dynamical system $\mathcal C$ to a combinatorial object  $\mathcal D$.

For $z \in Z$, the cell $[z] = \mathcal A^{-1}(z) \subset M$. If the cells are disjoint, the collection $K = \{[z]|~z \in Z\}$ is  called a partition of the state space $M$; whereas,  if a pair $[z] \cap [z'] \neq \emptyset$, the collection is called a cover.   

An abstraction is an over-approximation if for any $(t,x) \in \mathbb R_{\geq 0} \times M$
\[\mathcal A \circ \phi(t,x) \subseteq \Phi(t, \mathcal A(x)); \] 
$\mathcal A$ is an under-approximation if  
\[\Phi(t, \mathcal A(x))  \subseteq \mathcal A \circ \phi(t,x). \]
If $\mathcal A$ is a both under- and an over-approximation, then it is called a complete abstraction. For the questions related to safety, one might choose an over-approximation; whereas, for the questions corresponding to reachability, one might work with an under-approximation. Conservativeness of the abstraction,  say over-approximation, is measured by the volume,
\[\sup_{t \in \mathbb R_{\geq 0}} \max_{z \in Z} \mathrm{vol} (\Phi(t, z) \setminus \mathcal A \circ \phi(t,[z])).  \] 

%We will also discuss the following abstraction
%$$\Phi: Z $$

Below, we sketch a number of examples discussed during the talk. 

%\fbox{Add in the examples references to the works which have the same flavour as discussed here.}

\begin{example}
Suppose $\{U_z | z \in Z\}$ is a finite family of subsets covering $M$. Let $\mathcal D$ be given by $Z$ and $\Phi(t,z) =  \mathcal A \circ  \phi(t,[z])$.
Pick an order on Z. We define the abstraction $\mathcal A$ by 
\begin{equation} \label{TrivialAbstraction} \mathcal A: x \mapsto \min\{z \in Z |~x \in U_z \}.\end{equation}
%\[ \mathcal A: x \mapsto \{k \in Z |~x \in U_k \}.\]
As a consequence of the definition of $\Phi$, the abstraction $\mathcal A$ is an over-approximation. In this example, the computation of $\Phi$ might be tedious if not impossible. Therefore, an approximation is in place. 

To this end, we define
$$\mathrm{pol}\{v_1,...v_l\} = \left\{\sum_{i=1}^l \alpha_i v_i(x) \left |~\alpha_i \geq 0 \hbox{ and } \sum_{i=1}^l \alpha_i^2 = 1\right.\right\}.$$
Let $L = \{L_i|~i = 1, \hdots, l\}$ be a family of linear vector fields, and define multivalued map $F(x) = \mathrm{pol} L(x)$. Suppose that $\xi \in F(x)$, and define
\[\Phi(t,z) = \mathcal A \circ \mathrm{pol}\{ \phi^{L_1}(t,[z]), \hdots, \phi^{L_l}(t,[z]) \}.\]   
The over-approximation might be relatively conservative, but the computation is simplified as the flow maps are linear in the second argument. The algorithm can be additionally simplified if the sets $U_{k}$ are polyhedral (in local patches).
\end{example}

\begin{example}
Suppose that there exists a Finsler-Lyapunov (smooth) function \cite{ForliSepulchre} $V: TM \to \mathbb R$ (where $\pi: TM \to M$ is the tangent bundle) such that
\begin{enumerate}
\item $V(v) > 0$ for all $v \in TM \setminus 0_M$.
\item There is $p \in \mathbb N$ such that $V(\lambda v) = \lambda^p V(v)$ for all $v \in TM$ and $\lambda > 0$. 
\item There is  $p \in \mathbb N$ such that  $V(v+w)^{\frac{1}{p}} < V(v)^{\frac{1}{p}} + V(w)^{\frac{1}{p}}$ for all $v, w \in TM$ with $\pi(v) = \pi(w)$.
\end{enumerate}
The function $V$ defines metric $\rho$ on $M$ \cite{Tamassy2008483}
\[\rho(x_1,x_2) = \inf_{\gamma \in \Gamma(x_1,x_2)} \int_{I} V(\dot \gamma)^{\frac{1}{p}} ds,\]
where $I = [0, 1]$, $\dot \gamma = \gamma_*(d/dt)$, $\Gamma(x_1,x_2)$ is the set of curves $I \to M$ with $\gamma(0) = x_1$ and $\gamma(1) = x_2$.
Following Theorem~1 in [Forni and Sepulchre], if $dV: TM \to T^*(TM) $ satisfies the following inequality written in local coordinates
\[DV(x,w)(\xi(x), D\xi(x)w) \leq - \alpha (V(x,w)),~ \hbox{ for all } (x,v) \in TM.\]
where $\alpha$ is a non-decreasing continuous function. 
Then $\rho(\phi(t,x_1), \phi(t,x_2)) \leq \alpha (\rho(x_1,x_2))$. Hence, the system incrementally stable \cite{Angeli02}.

Since the state space $M$ is compact, it is possible to cover $M$ by the finite family $\{D(x_z,r_z) |~z \in Z\}$ of disks $D(x,r) = \{y \in M |~\rho(x,y) < r \}$ \cite{Frehse2008}.  
We define the abstraction $\mathcal A$ as in~\eqref{TrivialAbstraction}, and the combinatorial system $\mathcal D$ by $Z$ and $\Phi(t,z) = \mathcal A \phi(t,x_z)$. 
The abstraction $\mathcal A$ is an over-approximation. We note that computation of $\Phi$ amounts to simulating the dynamical system $\mathcal C$ for a finite number of initial conditions $x_z$.

%In this example, we relax the concept of over-approximation by requiring that for any $t$ there is a sequence $\{t_n\}$ such that 
%\[\bigcup_{\{t_n\}}\mathcal A \phi(t_n,x) \subseteq \Phi(t, \mathcal A(x)); \] 
% 
\end{example}

%%%%%%%%%
\begin{example}
Let $\xi$ be a Morse-Smale vector field on $M$ [Palis and de Melo]. Recall, a vector field $\xi \in \frak X^r(M)$ will be called Morse-Smale provided it satisfies
the following five conditions:
\begin{enumerate}
\item $\xi$ has a finite number of singular points, say $\beta_1,...,\beta_k$, each
hyperbolic, \item $\xi$ has a finite number of closed orbits (periodic solutions), say
$\beta_{k+1},...,\beta_N$, each hyperbolic; \item For any $x \in M,~\alpha(x) = \beta_i$
and $\omega(x) = \beta_j$ for some $i$ and $j$;
%\item If $\beta_i$ is a closed orbit then
%there is no $x \in M-\beta_i$ such that both $\alpha(x) = \beta_i$ and $\omega (x) = \beta_i$.
 \item $\Omega(\xi) = \{\beta_1,...,\beta_N \}$; 
 \item The stable and unstable manifolds associated with the $\beta_i$ have transversal
intersection.
\end{enumerate}
%The set of all Morse-Smale vector fields on $M$ is denoted by $\frak S^r(M)$.

The sets $\beta_1,...,\beta_N$ will be called the singular elements of the vector field
$\xi$. The set of the singular elements of $\xi$ will be denoted by ${\cal C}r(\xi)$. The stable  (unstable) manifold of $\xi$ at a singular element $\beta$ is denoted by $ W^s(\beta_i)$ ($W^u(\beta_i)$).

We define a partial order relation on the singular elements
of a Morse-Smale vector field: $\beta_i \succ \beta_j$ will mean that $W(\beta_i,\beta_j) \equiv W^u(\beta_i) \cap W^s(\beta_j) \neq \emptyset$.
%there is a trajectory not equal to $\beta_i$ nor $\beta_j$ whose $\alpha$-limit set is
%$\beta_i$ and whose $\omega$-limit set is $\beta_j$. Then we take
%\[Z = Cr(\xi) \cup \bigcup_{p }\]

%Then $\succ$ satisfies:
%\begin{enumerate}
%\item It is never true that $\beta_j \succ \beta_i$; \item If $\beta_i \succ \beta_j$ and
%$\beta_j \succ \beta_l$ then $\beta_i \succ \beta_l$; \item If $\beta_i \succ \beta_j$
%then $\dim (W_{\beta_i}^u) \geq \dim (W_{\beta_j}^u)$ and equality can only happen if
%$\beta_j$ is a closed orbit.
%\end{enumerate}

Consequently, each $W(\beta_i, \beta_j)$ is a cell, with the property that if $x \in W(\beta_i, \beta_j)$ then $\phi(t,x) \in W(\beta_i, \beta_j)$ for all $t \in \mathbb R$. Since the number of singular elements is finite, we can define $\mathcal D$ by 
\[Z = \{W(\beta_i,\beta_j)|~\beta_i \succ \beta_j \} \hbox{ and } \Phi(t,z) = z. \]

\end{example}

%%%%%%%%%%%%%%%%%%%%%%%%%%%%%%%%%%%%%%%%%%%%
\begin{example}
On the state space $M$, we define a family of functions $\{V_i: M \to \mathbb R|~i = 1, \hdots, l\}$ that satisfy 
\begin{enumerate}
\item $d V_i(\xi)(x) \leq 0$.
%\item For any $V_i$ and any regular value $a$, there is $b \in \mathbb R$ such that  $V_i^{-1}(a) \subset (d V_i (\xi)) ^{-1}(b)$.
\item Let $\mathrm{Reg}(V_i)$ be the set of regular values of $V_i$. For any singular element $\beta$ of $\xi$, 
   \begin{itemize}
   \item if $V_i^{-1}(\mathrm{Reg}(V_i)) \cap W^s(\beta) \neq \emptyset$ then $W^u(\beta) \subset V_i^{-1}(V_i(\beta))$;
    \item if $V_i^{-1}(\mathrm{Reg}(V_i)) \cap W^u(\beta) \neq \emptyset$ then $W^s(\beta) \subset V_i^{-1}(V_i(\beta))$.
   \end{itemize}
\end{enumerate}

For each function $V_i$, we associate a family of regular values $A^i \equiv \{a^i_0, \hdots, a^i_k |~a^i_{k-1} < a^i_k\} \subset \mathbb R \cup \{-\infty, + \infty\}$. For $a^i_j \in A^i$, we define a shift operator $\sigma \equiv \sigma^i:  a^i_j \mapsto a^i_{j-1}$
We use the notation $z = (z_1, \hdots, z_l)$ and define a cells $[z]$ with $z_i \in A^i$ by  
$$[z] = \bigcap V_i^{-1}([\sigma z_{i}, z_i])$$

Let $\mathbb R_{\infty} \equiv \mathbb R \cup \{-\infty,+\infty\}$. For each $z \in  Z \equiv  A^1 \times \hdots \times A^l$, we define a cube $\Box_{z} \equiv [\underline b_{z_1} \overline b_{z_1}] \times \hdots \times [\underline b_{z_l} \overline b_{z_l}] \subset \mathbb R_{\infty}^l$ with $\underline b_{z_i}$ ($\overline b_{z_i}$) being  the minimal (maximal) time over the trajectories staring at $V_i^{-1}(\sigma z_i)$ and leaving  $V_i^{-1}(z_i)$ (If $V_i^{-1}([\sigma z_i, z_i])$ is a positive invariant set, this time is set to $+\infty$). We denote the set of cubes in $\mathbb R^l$ by $\mathrm{Box}$. As a consequence, the combinatorial system is characterised by a map 
$\Box: Z \to \mathrm{Box}$ defined by $z \mapsto \Box_z$.

The following operator $L$ will be instrumental: $L = (L_1, \hdots, L_l) \to \mathbb R_{\infty}^l$, where $L_i = \partial \circ \pi_i$,  $\pi_i$ is the projection on the $i$th component, and $\partial [\underline b, \overline b] = \overline b - \underline b$. 

We define, a combinatorial system $\mathcal D$ by $Z$ and $\Phi$ as  
\begin{eqnarray*}
\Phi(t,z) &=& \max \{z' \in A^1 \times \hdots \times A^l |~
\Box_z \equiv \Box_{z^0} < \Box_{z^1} \hdots < \Box_{z^m} \equiv \Box_{z'},~ \\
&~& L(\Box_{z^0} + \hdots + \Box_{z^m}) \leq (t, \hdots, t), \hbox{ and } z^{i-1} = \sigma z^i \hbox{ for } i = 1, \hdots ,m\}.
\end{eqnarray*}

By \cite{Sloth201380}, this abstraction is complete.
\end{example}

\bibliographystyle{eptcs}
\bibliography{AbstractionsLit}

\end{document}